\documentstyle[aps,prd,preprint]{revtex}
\begin{document}
\title{Self-Dual Model Coupled to Bosons}
\author{M. A. Anacleto\footnote{maa@fisica.ufpb.br}, J. R. S. Nascimento\footnote{jroberto@fisica.ufpb.br} and R. F. Ribeiro\footnote{rfreire@fisica.ufpb.br}}
\address{Departamento de F\' \i sica, Universidade Federal da Para\'\i ba, Caixa Postal 5008, 58051-970 Jo\~ao Pessoa, Para\'\i ba, Brazil}
\date{\today}
\maketitle
\begin{abstract}
In this paper we investigated the dynamics, at the quantum level,
 of the self-dual field minimally coupled to bosons. In this investigation
 we use the Dirac bracket quantization procedure to quantize the model. 
Also, the relativistic invariance is tested in connection with the elastic 
boson-boson scattering amplitude.
\end{abstract}

\newpage

The quantum field theory in (2 + 1) dimensions has been provided for 
explaing the quantum Hall effect and the high temperature superconductivity. 
Besides, interesting aspects like exotic statistics, fractionary spin 
and insights about existence of massive gauge field is present in the 
context of three dimensions models\cite{r}.

Of the gauge theories, the self-dual theories deserve special attention.
Self duality refers to theories in which the interactions have particular
 forms and special strengths such that the equations of motion reduce 
from second to first order differential equations. In the context of the 
electrodynamics self-dual (SD) model, originally proposed by 
Townsend, Pilch and 
Van Nieuwenhuizen~\cite{r1} has been the object of 
several investigations. In \cite{r2,r3} it was demonstrated the equivalence,
 on semiclassical level, between free SD with Maxwell Chern-Simons (MCS). 
This equivalence was also observed in the level of the Green 
functions \cite{r4}.

The exact equivalence between the self-dual model minimally coupled 
to a Dirac field and the MCS model with non-minimal magnetic 
coupling to fermions has been studied by Gomes et al\cite{r5}.
Canonical quantization of the self-dual model coupled to fermions has been
 studied by Girotti~\cite{r6}. In this study, one has observed that two new 
interactions terms arise, which are local in space and time and are
non-renormalizable by power counting. Relativistic invariance is 
tested in connection with the elastic fermion-fermion scattering amplitude.

In this paper we will study the self-dual model coupled to boson where 
we are using the Dirac bracket quantizantion procedure to quantize it. 
Lorentz invariance is tested in connection with the elastic boson-boson 
scattering amplitude. In this case, we demonstrate that the combined action 
of the non-convariant pieces that make up the interaction in the Hamiltonian, 
can be replaced by the minimal covariant field-current interaction.

We adopt the following Heaviside-Lorentz units, and put $\hbar=c=1$. The
metric tensor is $g^{\mu\nu} = diag(1,-1,-1)$ and antisymmetric tensor 
$\epsilon^{\mu\nu\rho}$ is normalized as $\epsilon^{012}=1$. Also, we have 
considered $\epsilon^{ij}=\epsilon^{0ij}$.  

The Lagrangian that describe the self-dual field coupled to bosons is 
written as
\begin{equation}
{\mathcal{L}}=-\frac{1}{2m}\epsilon^{\mu\nu\rho}(\partial_{\mu}f_{\nu})f_{\rho}
+\frac{1}{2}f^{\mu}f_{\mu}+(D_{\mu}\phi)^{*}(D^{\mu}\phi)
-M^{2}\phi^{*}\phi
\end{equation}
where the covariante derivative is given by 
$D_{\mu}=\partial_{\mu}
+\frac{ig}{m}f_{\mu}$, $f_{\mu}$ is the self-dual field and $\phi$ is the 
charged scalar field.

The momenta canonically conjugated  are 
\begin{equation}
\pi_{\alpha}=\frac{\partial\mathcal{L}}{\partial(\partial_{0}f_{\alpha})}
=-\frac{1}{2m}\epsilon_{0\alpha\rho}f^{\rho},
\end{equation}
\begin{equation}
\Pi=\frac{\partial\mathcal{L}}{\partial(\partial_{0}\phi)}=
\partial^{0}\phi^{\ast}-\frac{ig}{m}f^{0}\phi^{\ast},
\end{equation}
\begin{equation}
\Pi^{*}=\frac{\partial\mathcal{L}}{\partial(\partial_{0}
\phi^{\ast})}=\partial^{0}\phi+\frac{ig}{m}f^{0}\phi.
\end{equation}

The primary constraints are
\begin{equation}
P_{0}=\pi_{0} \approx 0,
\end{equation}
\begin{equation}
P_{i}=\pi_{i}+\frac{1}{2m}\epsilon_{ij}f^{j}\approx 0,
\end{equation}
where the sign of weak equality $(\approx)$ is used in the sense of Dirac 
\cite{r7,r8}. The canonical Hamiltonian density is given by
\begin{equation}
{\mathcal{H}}=\pi_{i}\partial^{0}f^{i}+\Pi\partial^{0}\phi+
\Pi^{*}\partial^{0}\phi^{*}-\mathcal{L},
\end{equation}
then,
\begin{eqnarray}
{\mathcal{H}}&=&\Pi^{*}\Pi-\partial_k\phi^{*}\partial^k\phi
+M^2\phi^{*}\phi+\frac{ig}{m}(f_k\phi^{*}\partial^k\phi
-\partial_k\phi^{*}f^{k}\phi)
\nonumber\\
&-&\frac{g^2}{m^2}f_kf^k\phi^{*}\phi
-\frac{1}{2}f^{\mu}f_{\mu}
+f^0\left[\frac{1}{m}\epsilon_{ij}f^{0}\partial^{i}f^{j}
-\frac{ig}{m}(\Pi\phi-\Pi^{*}\phi^{*})\right].
\end{eqnarray}

The primary Halmitonian is 
\begin{equation}
H_{P}=\int d^{2}x({\mathcal{H}}+U^{0}P_{0}+U^{i}P_{i}).
\end{equation}
where $U^{0}$ and $U^{i}$ are the Lagrange multipliers.

Imposing the consistency conditions to the constraint  
\begin{equation}
\dot{P_{0}}=\{P_{0},H_{P}\}_{P}=\{\pi_{0}(\vec{x}),H_{P}(\vec{y})\}_{P}
\approx0,
\end{equation}
we find the secondary constraint
\begin{equation}
\label{1}
S=f^{0}-\frac{1}{m}\epsilon_{ij}\partial^{i}f^{j}
+\frac{ig}{m}(\Pi\phi-\Pi^{*}\phi^{*})\approx0. 
\end{equation}

Imposing again the same condition of consistency to the constraints $P_{i}$ 
and S we can verify that no more constraint arise. So we can determine 
the Lagrange multipliers and all constraints are second class. Following 
the Dirac bracket quantization procedure we get the commutation relation  
in equal time of the dynamics variables
\begin{equation}
\label{2}
[f^{0}(\vec{x}),f^{j}(\vec{y})]=i\partial^{j}_{x}\delta(\vec{x}-\vec{y}),
\end{equation}
\begin{equation}
[f^{k}(\vec{x}),f^{j}(\vec{y})]=-im\epsilon^{kj}\delta(\vec{x}-\vec{y}),
\end{equation}
\begin{equation}
[f^{0}(\vec{x}),\pi_{k}(\vec{y})]=-\frac{i}{2m}\epsilon_{kj}
\partial^{j}_{x}\delta(\vec x-\vec y),
\end{equation}
\begin{equation}
[f^{j}(\vec{x}),\pi_{k}(\vec{y})]=\frac{i}{2}g_{k}^{j}
\delta(\vec{x}-\vec{y}),
\end{equation}
\begin{equation}
[\pi_{j}(\vec{x}),\pi_{k}(\vec{y})]=-\frac{i}{4m}\epsilon_{jk}
\delta(\vec{x}-\vec{y}),
\end{equation}
\begin{equation}
[f^{0}(\vec{x}),\phi(\vec{y})]=-\frac{g}{m}\phi(\vec{x})
\delta(\vec{x}-\vec{y}),
\end{equation}
\begin{equation}
[f^{0}(\vec{x}),\phi^{\dagger}(\vec{y})]=\frac{g}{m}\phi^{\dagger}(\vec{x})
\delta(\vec{x}-\vec{y}),
\end{equation}
\begin{equation}
[\phi(\vec{x}),\Pi(\vec{y})]=i\delta(\vec{x}-\vec{y}),
\end{equation}
\begin{equation}
\label{3}
[\phi^{\dagger}(\vec{x}),\Pi^{\dagger}(\vec{y})]=i\delta(\vec{x}-\vec{y}).
\end{equation}
and all other commutators are vanish.

The Hamiltonian that describe the quantum dynamics of the system is written as
\begin{eqnarray}
H&=&\int{d^{2}x}\left[\Pi^{\dagger}\Pi-\partial_k\phi^{\dagger}\partial^k\phi
+M^2\phi^{\dagger}\phi\right.
\nonumber\\
&+&\left.\frac{ig}{m}(f^{k}\phi^{\dagger}\partial_{k}\phi
-\partial_k\phi^{\dagger}f^k\phi)-\frac{g^2}{m^2}f_{k}f^{k}\phi^{\dagger}\phi
+\frac{1}{2}f^{0}f^{0}+\frac{1}{2}f^{i}f^{i}\right].
\end{eqnarray}
where we have considered the Wick order of the operators. We can simplify the
Hamiltonian eliminating the operator $f^0$ using the condition that is
 described by the Eq.(\ref{1}) so that it takes the form

\begin{equation}
H^{I}=H^{I}_{0}+H^{I}_{int},
\end{equation}
where
\begin{eqnarray}
H^{I}_{0}&=&\int{d^{2}x}\left[\frac{1}{2m^2}\epsilon^{ij}\epsilon^{kl}
(\partial_{i}f^{I}_{j})(\partial_{k}f^{I}_{l})+\frac{1}{2}f^{I}_{i}f^{I}_{i}
\right]
\nonumber\\
&+&\int{d^{2}x}\left[\Pi^{I\dagger}\Pi^{I}
-\partial_{k}\phi^{I\dagger}\partial^{k}\phi^{I}
+M^{2}\phi^{I\dagger}\phi^{I}\right],
\end{eqnarray}
and
\begin{eqnarray}
\label{5}
H^{I}_{int}&=&\int{d^2{x}}\left[\frac{ig}{m}(f^{Ik}\phi^{I\dagger}
\partial_{k}\phi^{I}
-\partial_{k}\phi^{I\dagger}f^{Ik}\phi^{I})\right.
\nonumber\\
&-&\left.\frac{g^2}{m^2}f^{I}_{k}f^{Ik}\phi^{I\dagger}\phi^{I}
-\frac{ig}{m^2}\epsilon^{kl}\partial_{k}f^{I}_l(\Pi^{I}\phi^{I}
-\Pi^{I\dagger}\phi^{I\dagger})\right.
\nonumber\\
&-&\left.\frac{g^2}{2m^2}(\Pi^{I}\phi^{I}-\Pi^{I\dagger}\phi^{I\dagger})
(\Pi^{I}\phi^{I}-\Pi^{I\dagger}\phi^{I\dagger})\right]. 
\end{eqnarray}
The superscript $I$ denotes field operators belonging to the interaction 
picture.

The rules of commutations relations in equal times obeyed by operators of 
field in interaction pictures are exactly the equations (\ref{2})-(\ref{3}). 
The motion equations that satisfy the operators
 $\phi^{I}$ e $\phi^{I\dagger}$ are,
\begin{equation}
\label{6}
\partial_{0}\phi^{I}=i[H^{I}_{0},\phi^{I}]=\Pi^{\dagger},
\end{equation}
\begin{equation}
\label{7}
\partial_{0}\phi^{I\dagger}=i[H^{I}_{0},\phi^{I\dagger}]=\Pi.
\end{equation}
and the correspondent propagator of bosons $\Delta(p)$ in the momentum space is
\begin{equation}
\Delta(p)=\frac{i}{p^{2}-M^{2}+i\epsilon} .
\end{equation}
The Feynman propagator of the self-dual field $f^{I}_{i}, i=1,2$ 
is given by
\begin{equation}
D_{lj}(k)=\frac{i}{k^{2}-m^{2}+i\epsilon}(-m^{2}g_{lj}+k_{l}k_{j}
-im\epsilon_{lj}k_{0})=D_{jl}(-k)
\end{equation}
as has been obtained by Girotti\cite{r6}.

Finally, the Hamiltonian of interactions described in terms of the fundamental
fields is written as
\begin{eqnarray}
\label{8}
H^{I}_{int}&=&\int{d^2{x}}\left[\frac{ig}{m}(f^{Ik}\phi^{I\dagger}
\partial_{k}\phi^{I}-\partial_{k}\phi^{I\dagger}f^{Ik}\phi^{I})\right.
\nonumber\\
&-&\left.\frac{g^2}{m^2}f^{I}_{k}f^{Ik}\phi^{I\dagger}\phi^{I}
-\frac{ig}{m^2}\epsilon^{kl}\partial_{k}f^{I}_l(\partial_{0}\phi^{I\dagger}
\phi^{I}-\phi^{I\dagger}\partial_{0}\phi^{I})\right.
\nonumber\\
&-&\left.\frac{g^2}{2m^2}(\partial_{0}\phi^{I\dagger}\phi^{I}
-\phi^{I\dagger}\partial_{0}\phi^{I})
(\partial_{0}\phi^{I\dagger}\phi^{I}-\phi^{I\dagger}
\partial_{0}\phi^{I})\right].
\end{eqnarray}
Observe that the equation (\ref{8}) contain four terms. The first 
term 
is the spatial part of the field-current interaction. The third term is the 
magnetic field interacting with temporal component of the current. Whereas 
the second term is the spatial part of the gauge-boson field 
interaction and the fourth term is the interaction of a kind of temporal 
components of currents. Unlike the case of MCS minimally coupled to bosons, 
these extra terms are strictly local in space-time. Also, they are 
non-renormalizable by power counting.   

The next step we will verify the Lorentz invariance of the theory. To do this, we will 
evaluate the contibuition of order $g^2$ to the lowest order   elastic 
boson-boson scattering amplitude which can be grouped into four different 
kind of terms,
\begin{equation}
\label{9}
S^{(2)}=\sum_{\alpha=1}^{4}S_{\alpha}^{(2)},
\end{equation}
where
\begin{eqnarray}
S_{1}^{(2)}&=&\frac{g^2}{2m^2}\int\int d^{3}xd^{3}y<\varphi_{f}|
T\{:\{\phi^{\dagger}(x)\partial_{j}\phi(x)-\partial_{j}\phi^{\dagger}(x)
\phi(x)\}f^{j}(x):
\nonumber\\
&\times&:\{\phi^{\dagger}(y)\partial_{l}\phi(y)
-\partial_{l}\phi^{\dagger}(y)\phi(y)\}f^{l}(y):\}|\varphi_{i}>,
%\end{eqnarray}
%\begin{eqnarray}
\\
S_{2}^{(2)}&=&-\frac{g^2}{m^3}\int\int d^{3}xd^{3}y<\varphi_{f}|
T\{:\{\phi^{\dagger}(x)\partial_{j}\phi(x)-\partial_{j}\phi^{\dagger}(x)
\phi(x)\}f^{j}(x):
\nonumber\\
&\times&:\{\epsilon_{il}\partial^{i}f^l(y)(\partial_{0}\phi^{\dagger}(y)
\phi(y)-\phi^{\dagger}(y)\partial_{0}\phi(y)\}:\}|\varphi_{i}>,
%\end{eqnarray}
%\begin{eqnarray}
\\
S_{3}^{(2)}&=&\frac{g^2}{2m^4}\int\int d^{3}xd^{3}y<\varphi_{f}|
T\{:\{\epsilon_{kj}\partial^{k}f^j(x)(\partial_{0}\phi^{\dagger}(x)
\phi(x)-\phi^{\dagger}(x)\partial_{0}\phi(x)\}:
\nonumber\\
&\times&:\{\epsilon_{il}\partial^{i}f^l(y)(\partial_{0}\phi^{\dagger}(y)
\phi(y)-\phi^{\dagger}(y)\partial_{0}\phi(y)\}:\}|\varphi_{i}>,
%\end{eqnarray}
%\begin{eqnarray}
\\
S_{4}^{(2)}&=&\frac{ig^2}{2m^2}\int\int d^{3}xd^{3}y\delta(x-y)<\varphi_{f}|
T\{:\{\partial_{0}\phi^{\dagger}(x)\phi(x)-\phi^{\dagger}(x)\partial_{0}\phi(x)\}:
\nonumber\\
&\times&:\{\partial_{0}\phi^{\dagger}(y)\phi(y)-\phi^{\dagger}(y)
\partial_{0}\phi(y)\}:\}|\varphi_{i}>.
\end{eqnarray}
Here, $T$ is the chronological ordering operator, whereas $|\varphi_{i}>$
and $<\varphi_{f}|$ denote the initial and final state of the reaction, 
respectively. For the case under analysis, both $|\varphi_{i}>$ and 
$<\varphi_{f}|$ are two-boson states.

In terms of the initial $(p_{1},p_{2})$ and the final momenta
$(p_{1}^{\prime},p_{2}^{\prime})$, the partial amplitude are
\begin{eqnarray}
\label{10}
S_{1}^{(2)}&=&-g^{2}N_{p}(2\pi)^{3}\delta^{3}(p_{1}^{\prime}+p_{2}^{\prime}
-p_{1}-p_{2})\{(p_{1}^{\prime}+p_{1})_{j}(p_{2}^{\prime}+p_{2})_{l}
\frac{1}{m^2}D^{jl}(k)+p_{1}\leftrightarrow p_{2}\}\nonumber\\
%\end{eqnarray}
%\begin{eqnarray}
\\
S_{2}^{(2)}&=&-g^{2}N_{p}(2\pi)^{3}\delta^{3}(p_{1}^{\prime}+p_{2}^{\prime}
-p_{1}-p_{2})\{(p_{1}^{\prime}+p_{1})_{j}(p_{2}^{\prime}+p_{2})_{0}
\frac{1}{m^2}\Gamma^{j}(k)
\nonumber\\
&+&(p_{1}^{\prime}+p_{1})_{0}(p_{2}^{\prime}+p_{2})_{j}
\frac{1}{m^2}\Gamma^{j}(-k)+p_{1}\leftrightarrow p_{2}\}
%\end{eqnarray}
%\begin{eqnarray}
\\
S_{3}^{(2)}&=&-g^{2}N_{p}(2\pi)^{3}\delta^{3}(p_{1}^{\prime}+p_{2}^{\prime}
-p_{1}-p_{2})\{(p_{1}^{\prime}+p_{1})_{0}(p_{2}^{\prime}+p_{2})_{0}
\frac{1}{m^2}\Lambda(k)+p_{1}\leftrightarrow p_{2}\}
%\end{eqnarray}
%\begin{eqnarray}
\\
\label{11}
S_{4}^{(2)}&=&-g^{2}N_{p}(2\pi)^{3}\delta^{3}(p_{1}^{\prime}+p_{2}^{\prime}
-p_{1}-p_{2})\{(p_{1}^{\prime}+p_{1})_{0}(p_{2}^{\prime}+p_{2})_{0}
\frac{i}{m^2}+p_{1}\leftrightarrow p_{2}\}
\end{eqnarray}
where
\begin{eqnarray}
\frac{1}{m^2}D^{jl}(k)&=&\frac{i}{k^{2}-m^{2}+i\epsilon}\left(-g^{jl}
+\frac{k^{j}k^{l}}{m^{2}}-\frac{i}{m}\epsilon^{jl}k_{0}\right),
%\end{eqnarray}
%\begin{eqnarray}
\nonumber\\
\frac{1}{m^2}\Gamma^{j}(k)&=&\frac{i}{k^{2}-m^{2}+i\epsilon}\left(\frac{i}{m}
\epsilon^{jl}k_{l}+\frac{k^{j}k^{0}}{m^{2}}\right),
%\end{eqnarray}
%\begin{eqnarray}
\nonumber\\
\frac{1}{m^2}\Lambda(k)&=&\frac{i}{k^{2}-m^{2}+i\epsilon}
\left(-\frac{k^{l}k_{l}}{m^{2}}\right),
\end{eqnarray}
\begin{equation}
k\equiv(p_{1}^{\prime}-p_{1})=(p_{2}^{\prime}-p_{2})
\end{equation}
is the momentum transfer.
Substituting the Eqs.(\ref{10})-(\ref{11}) into Eq.(\ref{9}) we find
\begin{equation}
\label{amp}
S^{(2)}=-g^{2}N_{p}(2\pi)^{3}\delta^{3}(p_{1}^{\prime}+p_{2}^{\prime}
-p_{1}-p_{2})\{(p_{1}^{\prime}+p_{1})_{\mu}(p_{2}^{\prime}+p_{2})_{\nu}
\frac{1}{m^2}D^{\mu\nu}(k)+p_{1}\leftrightarrow p_{2}\}   
\end{equation}
where
\begin{equation}
\label{propl}
\frac{1}{m^2}D^{\mu\nu}(k)=-\frac{i}{k^{2}-m^{2}+i\epsilon}\left(g^{\mu\nu}
-\frac{k^{\mu}k^{\nu}}{m^{2}}+\frac{i}{m}\epsilon^{\mu\nu\alpha}k_{\alpha}
\right),
\end{equation}
is the propagator of the self-dual field $f^{\mu}$. As can be seen
the amplitude $S^{(2)}$ is the scalar of Lorentz. Observe, also, that the 
theory
 has passed in the test of the relativistic invariance. On the other hand,
in the tree approximantion is allowed to replace all the non-covariant terms 
in $ H_{int}^I$, Eq.(\ref{8}), by the minimal covariant interaction 
$\frac{g}{m}J_{\mu}^If^{\mu}$. Where $J_{\mu}^I$ is given by
\begin{equation}
J_{\mu}^{I}=i(\phi^{I\ast}\partial_{\mu}\phi^{I}-\partial_{\mu}
\phi^{I\ast}\phi^{I})-\frac{g}{m}f^{I}_{\mu}\phi^{I\ast}\phi^{I}
\end{equation}
Observe that the high energy behavior of the propagator in 
Eq.(\ref{propl}) is radically different from the MCS theory in the Landau
gauge\cite{rj},
\begin{equation}
D^{\mu\nu}_{L}(k)=-\frac{i}{k^{2}-m^{2}+i\epsilon}\left(g^{\mu\nu}
-\frac{k^{\mu}k^{\nu}}{k^{2}}+\frac{im}{k^2}\epsilon^{\mu\nu\alpha}k_{\alpha}
\right),
\end{equation}
 and therefore, the self-dual model coupled to bosons is a 
non-renormalized theory as we have noted previously by power counting. 

Finally, we conclude in this paper that the 
SD model minimally coupled to bosons bears no resemblance with the 
renormalizable model defined by the MCS field minimally coupled to bosons. 
The equivalence between SD and MCS when coupled to bosons is under 
investigation.

\acknowledgements
We are grateful to F. A. Brito for useful discussions. This work was 
partially supported by Conselho Nacional de Desenvolvimento 
Cient\' \i fico e Tecnol\'ogico(CNPq) and Coordena\c c\~ao de 
Aperfei\c coamento de Pessoal de N\' \i vel Superior (Capes)

\end{document}